\documentclass[twocolumn,dvipsnames]{aastex631}
\usepackage{physics}
\usepackage[hyphens]{url}

\shorttitle{Inertial modes in the Sun}
\shortauthors{Triana et al.}

\begin{document}

\title{Identification of inertial modes in the solar convection zone}

\author[0000-0002-7679-3962]{Santiago A. Triana}
\affiliation{Royal Observatory of Belgium, Ringlaan 3, 1180 Brussels, Belgium}

\author[0000-0002-2671-8796]{Gustavo Guerrero}
\affiliation{Federal University of Minas Gerais, Av. Pres. Antônio Carlos, 6627, Belo Horizonte, MG,  31270-901, Brazil}
\affiliation{New Jersey Institute of Technology, Newark, NJ 07102 USA}

\author[0000-0001-5747-669X]{Ankit Barik}
\affiliation{Johns Hopkins University, 3400 N. Charles Street, Baltimore, MD 21210, USA}

\author[0000-0003-3151-6969]{J\'er\'emy Rekier}
\affiliation{Royal Observatory of Belgium, Ringlaan 3, 1180 Brussels, Belgium}
\affiliation{Department of Earth and Planetary Science, University of California, Berkeley, CA 94720, USA}

\begin{abstract}

The observation of global acoustic waves (p modes) in the Sun has been key to unveiling its internal structure and dynamics. A different kind of wave, known as sectoral Rossby modes, have been observed and identified, which potentially opens the door to probing internal processes that are inaccessible through p mode helioseismology. Yet another set of waves, appearing as retrograde-propagating, equatorially antisymmetric vorticity waves, have also been observed but their identification remained elusive. Here, through a numerical model implemented as an eigenvalue problem, we provide evidence supporting the identification of those waves as a class of inertial eigenmodes, distinct from the Rossby mode class, with radial velocities comparable to the horizontal ones deep in the convective zone, but still small compared to the horizontal velocities towards the surface. We also suggest that the signature of tesseral-like Rossby modes might be present in the recent observational data.

\end{abstract}

\keywords{Solar oscillations (1515) --- Helioseismic pulsations (708) --- Hydrodynamics (1963)}

\section{Introduction} \label{sec:intro}

The Coriolis force in any rotating fluid body, from planetary cores and atmospheres to stars, supports the presence of inertial waves. Rossby waves, common in the Earth's atmosphere, are a particular subset of inertial waves. In the astrophysical literature Rossby waves are referred to as \emph{r-modes}, while the more general class of inertial waves are known, a bit confusingly, as \emph{generalized r-modes} \citep{lockitch1999}. Rossby waves play a fundamental role in the emission of gravitational waves in neutron stars \citep{andersson1997}, they have a major influence on Earth's weather \citep{michel2011}, and may be even present in the fluid cores of terrestrial planets affecting their global rotation \citep{triana2021}.

Retrograde-propagating vorticity waves, \emph{symmetric} with respect to the equator, have been observed and identified a few years ago as Rossby waves in the Sun by \cite{Loptien+18} and further confirmed by \cite{Hathaway+21} and \cite{gizon2021}. This discovery is highly relevant since {the damping rate of} Rossby waves is sensitive to turbulent flows, as opposed to the acoustic (p) modes also present in the solar convective zone. Thus, the observation and careful characterization of solar Rossby waves might be instrumental for the understanding of the internal solar dynamics so far elusive for  traditional p mode helioseismology. 

More recently, \cite{Hanson+22} provided observational evidence of a distinct set of High-Frequency, Retrograde-propagating (HFR) vorticity waves, penetrating at least to 3\% of the solar radius, but this time with an \emph{antisymmetric} vorticity with respect to the equator. However, the identification of these waves was left as an open question. They do not seem to fit the classical Rossby wave dispersion relation that, contrastingly, worked so well for the symmetric-vorticity waves described by \cite{Loptien+18}.

In this work we provide numerical evidence supporting the identification of the HFR waves as a class of inertial waves, different from the Rossby mode class, that span the whole convective zone depth but with dominant horizontal flows near the outer regions. Our numerical model is based on a simplified physical description of the convection zone, providing the eigenvalues and eigenvectors associated with the inertial modes that it may support. 

Non-axisymmetric (i.e. with azimuthal wave number $m\neq 0$) inertial eigenmodes drift in longitude according to their phase speed, so in this work we refer to them as waves or modes interchangeably.

\section{A model for inertial eigenmodes}

\subsection{Main description}

A starting model for solar inertial oscillations can be built by representing the solar convection zone as a spherical shell filled with an homogeneous, incompressible, and viscous fluid. The inner radius $r_c$ of the shell corresponds to the outer radius of the radiative core, with a value $r_c\sim0.71R_\odot$ \citep{jcd1991}, $R_\odot$ being the solar radius. The flow velocity associated with the vorticity waves in the convection zone is much smaller than $\Omega_\odot r_c$, where $\Omega_\odot$ is {a representative value of the Sun's angular speed}, thus the flow velocity $\vb{u}$ can be described to a good approximation as a small perturbation to the uniformly rotating background flow by the \emph{linear} Navier-Stokes equation:
\begin{equation}
    \partial_t\vb{u}+2\vb{\hat z}\times\vb{u}=-\nabla p+E\nabla^2\vb{u}.
    \label{eq:NS}
\end{equation}
The variables in the preceding equation are rendered dimensionless by taking $R_\odot$ and $1/\Omega_\odot$ as the units for length and time, respectively. We use $\vb{\hat z}$ as the unit vector along the solar spin axis, and $p$ is the reduced pressure. We introduce also the Ekman number $E$ defined as
\begin{equation}
    E=\frac{\nu_\mathrm{eff}}{\Omega_\odot r_c^2},
    \label{eq:E}
\end{equation}
where $\nu_\mathrm{eff}$ is an \emph{effective} or turbulent eddy viscosity. The flow velocity $\vb{u}$ follows a time dependence described by
\begin{equation}
    \vb{u}(\vb{r},t) =\vb{u}_0(\vb{r})\text{e}^{\lambda t}+\mathrm{cc},
\end{equation}
where $\lambda = (\sigma/\Omega_\odot) - \mathrm{i} (\omega/\Omega_\odot)$ is a complex number whose real part $\sigma/\Omega_\odot$ corresponds to the dimensionless decay rate (or `damping' for short) and the imaginary part $\omega/\Omega_\odot$ to the dimensionless eigenfrequency. We add the complex conjugate (cc) to keep $\vb{u}$ real. Then, we write the velocity amplitude $\vb{u}_0$ in the poloidal-toroidal decomposition:
\begin{equation}
    \vb{u}_0=\nabla\times\nabla\times \left[ \mathcal{P}(\vb{r})\,\vb{r} \right]+\nabla\times \left[ \mathcal{T}(\vb{r})\,\vb{r} \right],
    \label{eq:u0}
\end{equation}
{which automatically satisfies the incompressible continuity equation $\nabla\cdot\vb{u}=0$.}
We use spherical harmonic expansions for the angular dependence of the scalar functions $\mathcal{P,\,T}$. For instance, we write the toroidal scalar function $\mathcal{T}(\vb{r})$ as
\begin{equation}
\mathcal{T}(r,\theta,\phi)=\sum_{l=1}^{l_\mathrm{max}}\sum_{m=-l}^l T_{lm}(r) Y_l^m(\theta,\phi),
\end{equation}
where $Y_l^m$ are the spherical harmonics, $T_{lm}(r)$ is a radial function, and $l_\mathrm{max}$ determines the angular truncation level. A completely analogous expression goes for $\mathcal{P}(\vb{r})$. {It is sometimes useful to further subdivide the poloidal component into a radial part $\mathcal{Q}$ and a consoidal part $\mathcal{S}$ (known also as `spheroidal'). Their spherical harmonic components are written, respectively,
\begin{equation}
    Q_{lm}=\frac{l(l+1)}{r}P_{lm}, \quad S_{lm}=\frac{1}{r}\partial_r(r\,P_{lm}).
\end{equation}
Note that the horizontal components of the flow velocity are related to both $\mathcal{S}$ and $\mathcal{T}$, while the radial vorticity is related only to $\mathcal{T}$.}

The spherical harmonic expansion leads to a fully decoupled problem in the azimuthal wave number $m$, allowing us to consider a single $m$ at a time. If $m\neq 0$, inertial eigenmodes are either retrograde or prograde, which is determined by the sign of their \emph{phase speed $\omega/m$}. Note that the retrograde inertial eigenmode spectrum does not coincide in general with the prograde one. 
{We specify stress-free boundary conditions at both boundaries by requiring
\begin{equation}
\begin{split}
   &\left. P_{lm}(r) \right\rvert_{r=r_b} =\left. P_{lm}''(r) \right\rvert_{r=r_b}=0,\\
   &\left. T_{lm}'(r)\right\rvert_{r=r_b} - \left. T_{lm}(r)/r \right\rvert_{r=r_b} = 0,
\end{split}
\end{equation}
where $r_b$ denotes the inner or the outer radius of the convective zone and the prime ($'$) symbol denotes the radial derivative.}

Our numerical scheme involves expansions of the radial functions $P_{lm}(r)$ and $T_{lm}(r)$ in terms of Chebyshev polynomials. This allows us to write the problem represented by Eq.~(\ref{eq:NS}) as a generalized eigenvalue problem. {For more details see Appendix A. Solutions can be classified according to their equatorial symmetry, see Appendix \ref{sec:apx2} for a brief discussion on the nomenclature.}

{The use of an incompressible and uniform density fluid in our model might seem too drastic a simplification. However, if the large background density gradient across the convection zone is addressed with the anelastic approximation approach, Eq.~(\ref{eq:NS}) would remain almost the same except with $\vb{u}$ replaced by $\rho\vb{u}$ (thus satisfying $\nabla\cdot\rho\vb{u}=0$, $\rho$ being the fluid density). If the radial displacements are small compared to the horizontal ones, then the difference would be mainly in the viscous term, which has in general only a weak impact on the modes' frequencies. In fact, even when compared with a fully compressible model which considers a realistic
solar density profile \citep[such as][see their Fig.~14]{bekki2022}, the results of our model differ at most by 15\% for $m=2$, and less than 2\% for $m=16$ Rossby mode frequencies. }

\subsection{Choosing a rotating frame}
In some works wave frequencies are reported with respect to a reference frame rotating at $\Omega_\odot/2\pi=456$ nHz, known as the Carrington frame \citep[e.g.][]{gizon2021, Hathaway+21}, or using a frame rotating at $\Omega_\odot/2\pi=453.1$ nHz \citep[e.g.][]{Loptien+18,Hanson+22}. Both rotation rates are close to the mean rotation rate of the solar equator \citep{larson2018}. A wave with azimuthal wave number $m$ and frequency $\omega_0$ observed in a frame rotating at $\Omega_0$ has a frequency $\omega_1=\omega_0+m(\Omega_0-\Omega_1)$ when observed from a frame rotating at $\Omega_1$. Throughout the present work we use the Carrington frame, except where indicated otherwise.

\section{Results}

\begin{figure}
\plotone{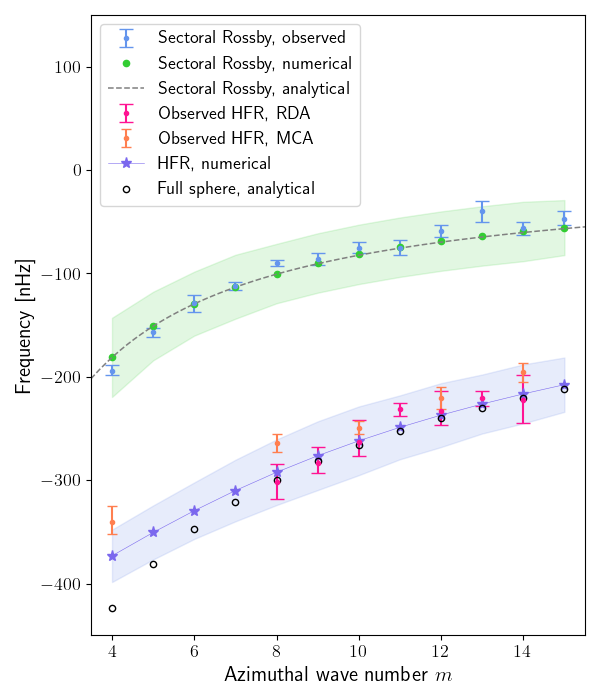}
\caption{Observed HFR vorticity wave frequencies {(relative to the Carrington frame)} from mode-coupling analysis (MCA, orange error bars) and from ring-diagram analysis (RDA, deep-pink error bars) compared to numerical inertial mode frequencies for $E=2.1\times10^{-4}$ and $\delta=0.29R_\odot$ (blue-violet line with stars). We include observed, numerical, and analytical sectoral Rossby mode frequencies {(as observed in a frame rotating at $\Omega_\odot/2\pi=453.1$ nHz)} for reference. Shaded areas represent the possible effect on the eigenfrequencies induced by the solar differential rotation. See main text for further details.
\label{fig:fig1}}
\end{figure}

The line width $\Gamma$ of the modes corresponds roughly to two times their decay rate constant $\sigma$, in analogy with a lightly damped harmonic oscillator. We have tuned the Ekman number so that the mean $[\sigma]$ of numerically computed modes (the HFR candidates), from $m=8$ to $m=14$, matches one half the mean $[\Gamma]$ of the observed line widths that were computed via ring-diagram analysis of Helioseismic and Magnetic Imager (HMI) measurements, as reported by \cite{Hanson+22} (see their Table 1, column 5). Thus, $[\sigma]\approx[\Gamma]/2=19.5$ nHz is obtained when $E=2.1\times10^{-4}$. The inertial mode frequency spectrum is {countably} dense, so there is always an inertial mode arbitrarily close to any frequency we might choose as a target for the eigensolver. {We look for lightly damped modes with simple spatial structure as they are easier to excite than modes with more complex structure}. The \emph{least} damped modes in a broad interval around the observed frequencies are plotted in Fig.~\ref{fig:fig1} (blue-violet stars labeled `HFR, numerical'). These modes are retrograde-propagating ($\omega<0$), have equatorially antisymmetric radial vorticity (i.e. equatorially symmetric flow velocity) and their dominant spherical harmonic component, $T_{lm}$, {near the outer boundary} is such that $l=m+1$, matching the corresponding characteristics of the observed HFR vorticity waves. We have extracted the observed HFR frequencies in Fig.~\ref{fig:fig1} (orange and deep pink error bars) from Table 1 and Table S1 of \cite{Hanson+22}, {transforming them to the Carrington reference frame}. For comparison, Fig.~\ref{fig:fig1} also presents the sectoral Rossby mode frequencies obtained with our model, which we used for validation, and the sectoral Rossby modes observed by \cite{Loptien+18}. {The shaded areas in Fig.~\ref{fig:fig1} represent a crude estimate of the effect of the solar differential rotation background on the eigenfrequencies, which we assume is comparable to the variation $\Delta\Omega=\Omega_\mathrm{max}-\Omega_\mathrm{min}$ of the background rotation rate $\Omega(r,\theta)$ but restricted to the $(r,\theta)$ region where a given mode has substantial amplitude. See Appendix \ref{sec:apx3}.}

\begin{figure*}
    \centering
    \includegraphics[width=0.805\paperwidth]{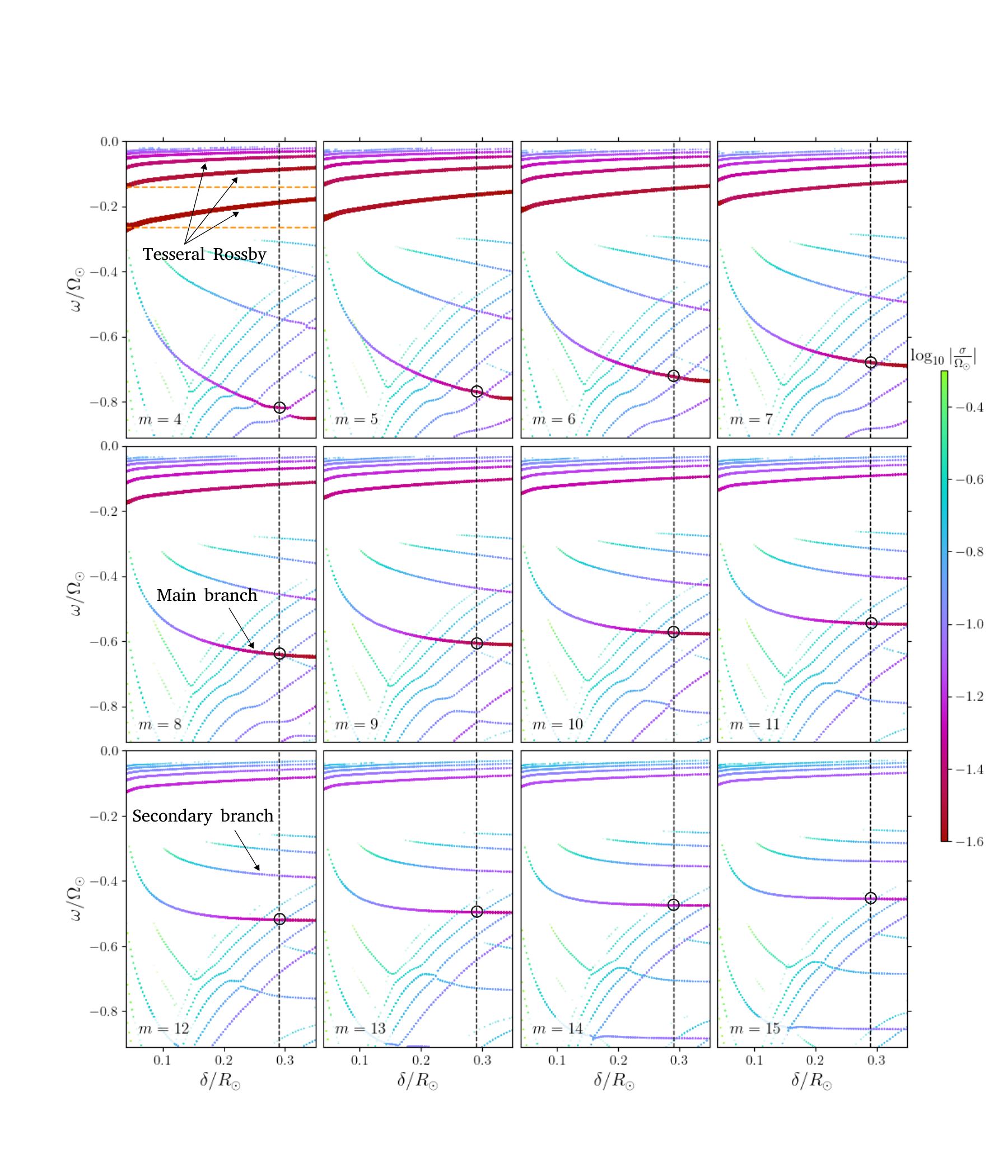}
    \caption{Eigenfrequencies of equatorially \emph{antisymmetric vorticity} modes as a function of the radial width $\delta$ of the convective zone for different azimuthal wave numbers $m$. Color and symbol size indicates the eigenmodes' damping decay factor $\sigma$ with larger deep-red points representing lightly damped modes (more likely to be excited), and smaller green-turquoise points representing heavily damped modes (less likely to be excited). The vertical black dashed line marks the Sun's actual convective zone width at $\delta=0.29R_\odot$. Black circles mark the modes we identify as the observed HFR waves. The horizontal orange-dashed lines in the $m=4$ panel are the frequencies of the $(l=5,m=4)$ and the $(l=7,m=4)$ tesseral Rossby modes according to Eq.~(\ref{eq:analytical}).}
    \label{fig:fig2}
\end{figure*}

With the exception of the purely toroidal inertial eigenmodes, the eigenfrequencies are sensitive to the radial width of the shell cavity. In Fig.~\ref{fig:fig2} we show the eigenfrequencies obtained with our numerical model by considering different convective zone widths $\delta$, and different azimuthal wave numbers $m$. Symbol color and size indicates $\log_{10}|\sigma/\Omega_\odot|$, with larger, deep red symbols indicating modes with little damping, which are easier to excite, and smaller, green-turquoise symbols indicating heavily damped modes. We see in the figure eigenmodes with small magnitude frequencies, typically with $|\omega/\Omega_\odot|<0.3$, particularly at low $m$, with very little damping. For instance, when $m=4$, the least damped mode has a frequency of about $\omega/\Omega_\odot\sim-0.2$ at $\delta=0.29R_\odot$ which increases in magnitude for smaller $\delta$. This class of modes have frequencies that generally decrease in magnitude and become more damped as $m$ increases. These are \emph{tesseral-like Rossby modes} whose frequencies are given to a very good approximation by the dispersion relation (\ref{eq:analytical}) but \emph{only} when the shell cavity is thin, i.e. when $\delta\lesssim 0.06R_\odot$. The top, left panel in Fig.~\ref{fig:fig2} shows the theoretical frequencies of the $(l=5,m=4)$ and the $(l=7,m=4)$ Rossby modes from Eq.~(\ref{eq:analytical}) as horizontal orange dash lines.

\begin{figure*}
\plottwo{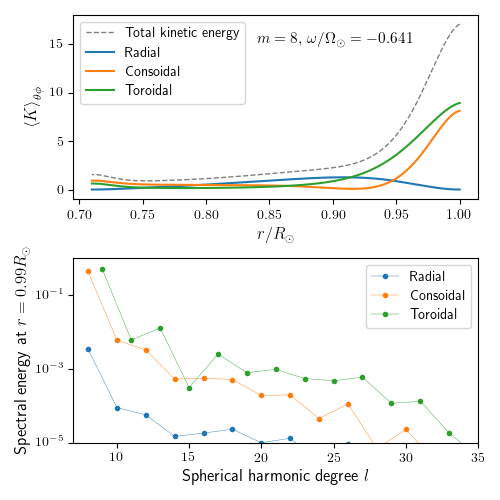}{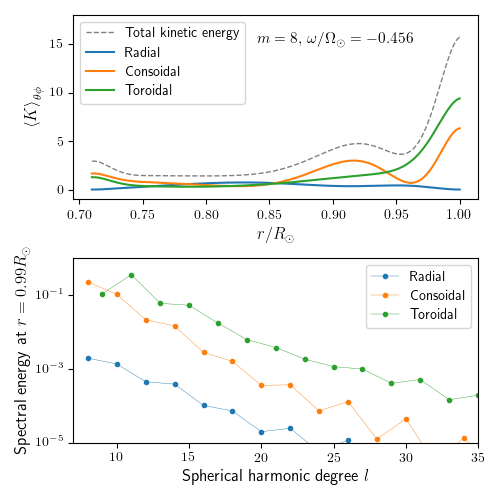}
\caption{Comparison of two $m=8$ modes, one on the main branch with $\omega/\Omega_\odot=-0.641$ (left column), and one on the secondary branch with $\omega/\Omega_\odot=-0.456$ (right column). Top panels show the radial profile of the kinetic energy associated with the radial, consoidal and toroidal components of the velocity (averaged over $\theta$ and $\phi$, modes are normalized to have unit kinetic energy). Bottom panels show the fractional energy content of each spherical harmonic component at $r=0.99R_\odot$ (fractions of the total kinetic energy at that radius). The convective zone width is $\delta=0.29R_\odot$ and the Ekman number is $E=2.1\times10^{-4}$. Note that the $(l,m)$ spherical harmonic component of the radial vorticity is proportional to $l(l+1) T_{l m}(r)/r$. } 
\label{fig:fig3}
\end{figure*}

We want to draw attention now to the modes with frequencies near $\omega/\Omega_\odot\sim -0.8$ for $m=4,\, \delta=0.29R_\odot$, progressively decreasing in magnitude as $m$ increases until reaching $\omega/\Omega_\odot\sim -0.45$ at $m=15,\,\delta=0.29R_\odot $. With the exception of tesseral-like Rossby modes mentioned earlier, these modes are the least damped in a wide frequency range as evidenced by their color and symbol size in Fig.~\ref{fig:fig2}. In the following we refer to these modes as the \emph{main branch}. Modes on this branch, and with $\delta=0.29R_\odot$, are marked with black circles. They appear also on Fig.~\ref{fig:fig1} labeled as `HFR, numerical'. A \emph{secondary} branch, with higher damping and lower frequencies in magnitude can also be identified. It corresponds to the branch with a mode at $\omega/\Omega_\odot\sim -0.55$ (for $m=4$, $\delta=0.29R_\odot$).  Again, as $m$ increases, their frequency progressively decreases in magnitude until reaching $\omega/\Omega_\odot\sim -0.35$ at $m=15$. Modes on the main branch have the particular property that their toroidal $(l=m+1,m)$ {and their consoidal $(l=m,m)$} components near the solar surface are dominant, matching the observed waves, while the modes on the secondary branch have a dominant toroidal $(l=m+3,m)$ component near the surface.  In Fig.~\ref{fig:fig3} we present a side-by-side comparison of the spectral amplitudes near the solar surface ($r=0.99R_{\odot}$) between two $m=8$ modes, one on the main branch (left column panels) and one on the secondary branch (right column panels) for $\delta=0.29R_\odot$.

Modes in the main branch appear as mainly toroidal from the surface, but they harbor a {non-negligible} poloidal component deep in the convective zone. The meridional cross sections in Fig.~\ref{fig:fig4} (top row) illustrate this point. Color indicates the dimensionless velocity amplitude for each spherical coordinate direction, and the amplitude of $\mathcal{P}$ and $\mathcal{T}$ (note that, as a result of an eigenvalue calculation, the overall amplitude of the eigensolution is arbitrary). Modes in the main branch appear to have no radial nodes in {the equatorial plane} while the modes in the secondary branch (not shown) appear to have one radial node. This suggests that the other weaker branches visible in Fig.~\ref{fig:fig2} represent branches with an increasing number of radial nodes.

Lastly, there is a correspondence between the modes in the main branch and a particular class of inertial modes of a full sphere. The inertial modes in a full sphere can be computed analytically \citep{greenspan1968,zhang2004}, and can be specified by three indices $(\nu,\mu,\kappa)$, following the notation used by \cite{greenspan1968}. The modes on the main branch reduce (when $\delta=R_\odot$) to the only retrograde eigenmodes of the full sphere with $\nu=m+2$ and $\kappa=m$. 
These analytical frequencies are plotted in Fig.~\ref{fig:fig1} as hollow black circles.

\section{Discussion}

The inclusion of turbulent viscous diffusion in our model give us the ability to discern which eigenmodes are more likely to be excited, although it does not tell us anything about the excitation mechanism itself. As explained earlier, we have tuned the Ekman number $E$ in order to match the mean damping rate of the $m=8,\ldots,14$ eigenmodes in the main branch with one-half of the mean line width of the observed HFR waves. This amounts to an effective viscosity $\nu_\mathrm{eff}\simeq 146\,\mathrm{km}^2/\mathrm{s}$. {Such relatively large viscosity prevents the appearance of thin internal shear layers that would otherwise appear at low viscosity, ultimately rendering the modes singular in the limit of vanishing viscosity \citep{rieutord2001}}. There is no direct way of measuring the value of the effective viscosity in the solar interior. However, observational and theoretical estimates of the turbulent magnetic diffusivity  at the solar surface find values of the order of $100\,\mathrm{km}^2/\mathrm{s}$ \citep[e.g.][]{abramenko+11,skokic2019,baumann+04}. Thus, by assuming that the turbulent magnetic Prandtl number is about unity \citep{kapyla2020}, our choice is entirely consistent. {Furthermore, we can combine the $30$ Mm length scale for the large scale convective flows as proposed by \cite{vasil2021} with the lower bound of about 10 m/s for the vertical convective velocity as estimated from observations by \cite{greer2016}, resulting in $\nu_\mathrm{eff}\simeq 300\,\mathrm{km}^2/\mathrm{s}$. Although within an order of magnitude of our original estimate, the foregoing calculation is admittedly oversimplified, especially considering that the effective viscosity may be anisotropic and scale dependent \citep{rincon2017}. }

For low $m$ values ($4 \leq m \leq 8$), the modes we find in the main branch are the second least damped after the tesseral-like Rossby mode family, while at higher $m$ the main branch becomes the least damped. As shown in Fig.~\ref{fig:fig1}, the observed HFR wave frequencies {as measured in the Carrington frame (rotating at 456 nHz) are a close match to the eigenfrequencies of the inertial modes in the main branch. Note that the sectoral Rossby mode frequencies shown in Fig.~\ref{fig:fig1} are relative to a frame rotating slightly slower at 453.1 nHz. If transformed to the Carrington frame, the sectoral Rossby modes computed from the dispersion relation Eq.~(\ref{eq:analytical}) would have frequencies systematically smaller than the observations (relative to that frame). The fact that theoretical HFR waves and sectoral Rossby waves both match well the observed frequencies but only when measured in two slightly different rotating frames is most likely a consequence of the different solar rotation rates as `sensed' by the two kinds of waves (see also Appendix \ref{sec:apx3}).}

Another important piece of evidence is provided by the spherical harmonic components. The eigenmodes in the main branch are characterized by their {large} $(l=m+1,m)$ spherical harmonic component of radial vorticity near the solar surface (see Fig.~\ref{fig:fig3}, left column panels), precisely like the observed HFR waves. {Note that the \emph{consoidal} component is also large compared to the radial component near the surface, but the radial vorticity observations are only sensitive to the toroidal component (see Appendix \ref{sec:apx2}). Although the toroidal and consoidal components are dominant near the surface, the radial component can be dominant in deeper regions, but never with as much amplitude as the other components near the surface.}

\begin{figure*}
\centering
\includegraphics[width=\linewidth]{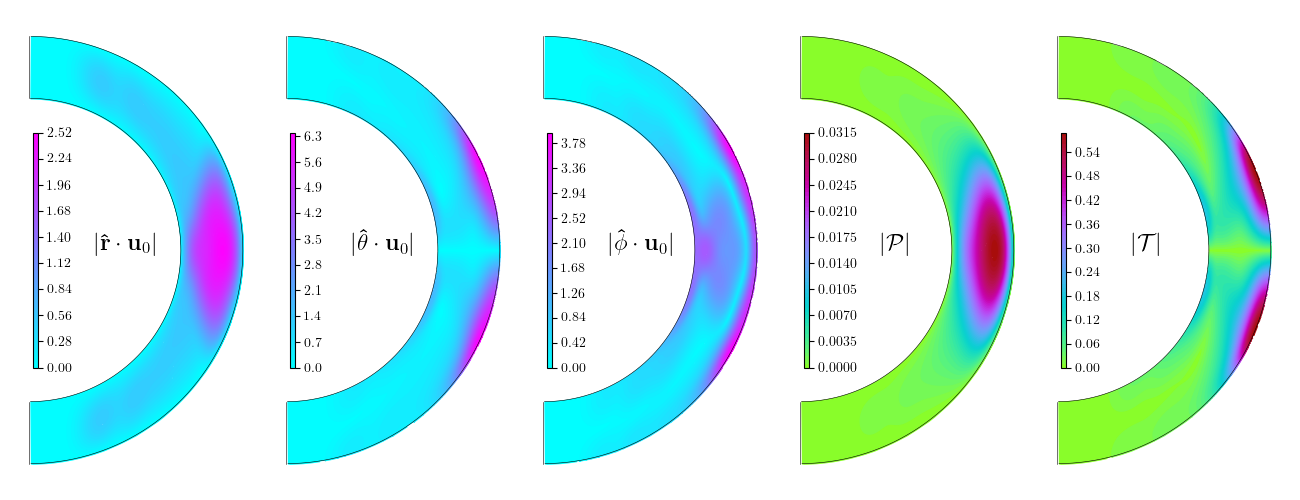}\\
\includegraphics[width=\linewidth]{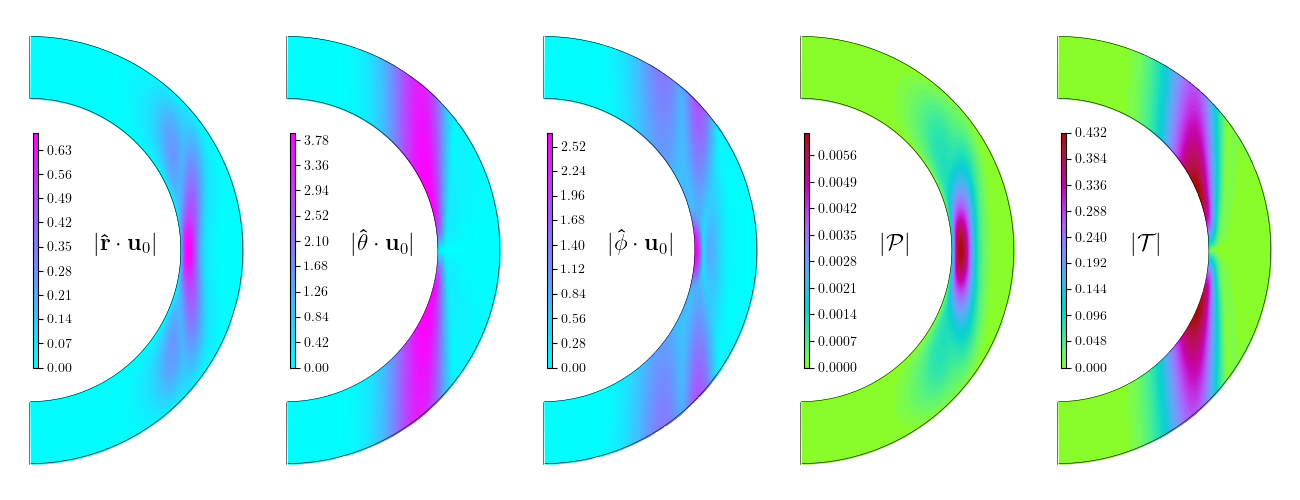}
\caption{Meridional cross sections of the $m=8$, $\omega/\Omega_\odot=-0.641$ eigenmode on the main branch {(top row), and the $m=8$, $\omega/\Omega_\odot=-0.118$ of the tesseral-like Rossby mode family (bottom row). This latter mode is essentially the same as the one reported by \cite{bekki2022}, see their Figure 11d. Note the absence of radial nodes of $\mathcal{P}$ or $|\vb{\hat r}\cdot\vb{u}_0|$ in the equatorial plane for the main branch mode. Ekman number here is $E=2.1\times10^{-4}$ and the convective zone width is $\delta=0.29R_\odot$.}} 
\label{fig:fig4}
\end{figure*}

It is tempting to adjust $\delta$ to match the observed frequencies but it cannot be done consistently for all $m$ numbers. A glance at Fig.~\ref{fig:fig2} reveals that modes with low $m$ numbers are much more sensitive to $\delta$ than modes with higher $m$. However, we cannot rule out entirely the possibility that modes with different $m$ numbers `perceive' a differently sized shell cavity, perhaps related in some way to the meridional circulation pattern in the solar convective zone. Nonetheless, we believe that accounting for the latitudinal differential rotation should be among the first refinements to be investigated.

It is interesting to note that the phase speed frequency $(\omega/2\pi)/m$ of the modes, in the range of 13 to $85\,\mathrm{nHz}$ in the retrograde direction, opens the possibility of \emph{co-rotation resonances} of the modes with the background latitudinal differential rotation \citep{baruteau2013,guenel2016}. In a reference frame co-rotating with the solar equator, the latitudinal differential rotation appears as retrograde azimuthal flow with an angular rotation rate decreasing continuously from 
$\simeq 120\,\mathrm{nHz}$ near the poles until vanishing at the equator. There is therefore a region in the convective zone in which the Doppler-shifted frequency vanishes for each mode. It is then conceivable that the modes draw their energy from the differential rotation. That being said, there is presumably some kind of interaction between the large-scale convective flow and the inertial waves with comparable time and length scales. In fact, at $m>15$ the waves appear indistinguishable from convection according to \cite{Loptien+18}. Thus, an adequate understanding of the interplay among differential rotation, waves, and large scale convection is desirable.   

The retrograde $(l=m+1,m)$ tesseral-like Rossby modes in our model have rather small damping rates, particularly at low $m$, which makes them candidates as well to be excited and observed. The power spectra presented by \cite{Hanson+22} (see their Fig.~1) shows indeed some power at low frequencies and low $m$ numbers in the $l=m+1$ channel, which might contain the signature of the tesseral-like Rossby modes. Although it is not clear if the observed power in that region is not caused by noise or some other artifact \citep[e.g. sunspots, see Appendix C of][]{gizon2021}. These tesseral-like Rossby modes have a ratio of toroidal to poloidal kinetic energy much larger than one (a distinguishing feature of \emph{all} Rossby modes), in contrast to the modes in the main branch. They are mostly columnar, with very little amplitude in the equatorial region as Fig.~\ref{fig:fig4} (bottom row) shows. According to our calculations, their spherical harmonic spectra at the surface shows significant contributions from the toroidal $l=m+1,\,m+3,\,m+5$ and higher components which matches, at least qualitatively, the data shown in Fig.~S4 of \cite{Hanson+22}.

Clearly, the dispersion relation (\ref{eq:analytical}) fails to hold for tesseral-like Rossby modes in a deep spherical shell, although it still holds for \emph{sectoral} Rossby modes, regardless of the shell depth. The latter are purely toroidal (i.e. purely horizontal fluid displacements) and their frequency or damping are essentially unaffected by the shell depth. The tesseral-like Rossby modes with symmetric vorticity, in addition to the sectoral ones, exhibit similar behaviour as the tesseral-like Rossby modes with antisymmetric vorticity. As demonstrated by the $m=4$ panel in Fig.~\ref{fig:fig2}, their frequency is no longer described by the dispersion relation (\ref{eq:analytical}) and they become increasingly damped as the shell depth increases. This might explain the apparent lack of symmetric vorticity modes other than sectoral Rossby in the analysis by \cite{Loptien+18}.

\begin{figure*}
\plottwo{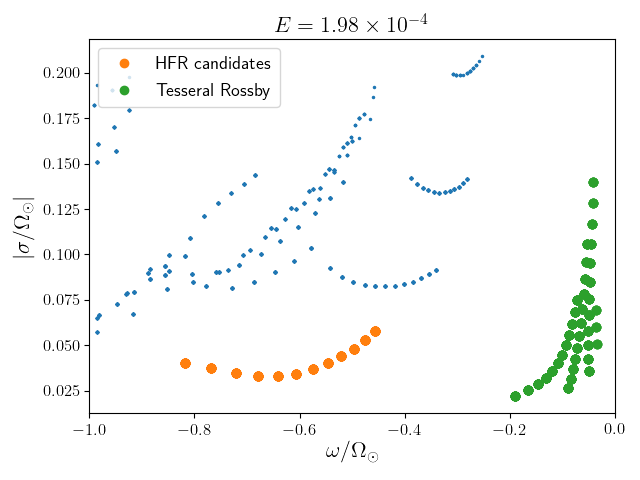}{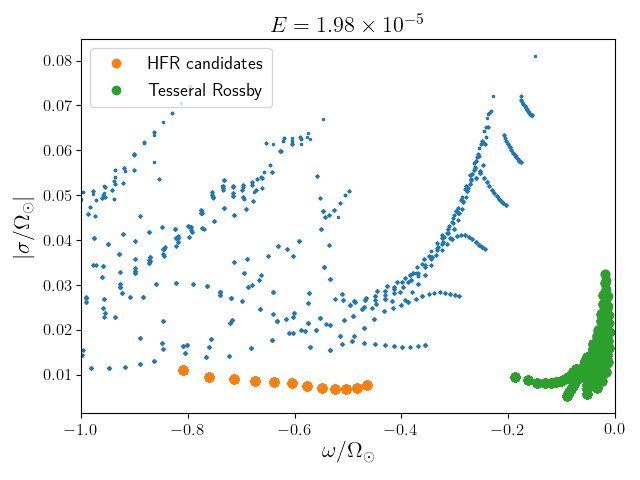}
\caption{Reducing the Ekman number by an order of magnitude has only a weak influence on the eigenfrequencies $\omega$ but a strong one on the damping rates $\sigma$. The plot on the left is computed for $E=1.98\times10^{-4}$ and the one on the right for $E=1.98\times10^{-5}$, both for $\delta=0.29R_\odot$, modes from $m=4$ to $m=15$ are included. Even at a reduced Ekman number the HFR modes (orange dots) remain as the least damped ones, along with some tesseral-like Rossby modes (green dots). Blue dots represent the remaining inertial modes.  
\label{fig:twoE}}
\end{figure*}

To conclude this section, we note that the Ekman number has a strong influence on the damping rates while only a weak influence on the eigenfrequencies. But even if the Ekman number is reduced by an order of magnitude as shown in Fig.~\ref{fig:twoE}, the tesseral-like Rossby modes and the modes on the main branch persist as the least damped ones, and the identification of the observed vorticity waves as modes on the main branch remains valid.

\section{Summary and conclusion}

We presented evidence supporting the identification of the high frequency retrograde (HFR) vorticity waves measured recently by \cite{Hanson+22} as a particular class of inertial modes of a deep spherical shell. Our findings are based on a relatively simple numerical model representing the solar convective zone as a homogeneous, incompressible and viscous fluid in a rotating spherical shell. The eigenmodes of this system correspond to oscillations where the Coriolis force is the restoring force. We recover the Rossby mode frequencies described by the well-known dispersion relation (\ref{eq:analytical}) but only for sectoral Rossby modes. We find that Rossby modes other than sectoral are also well described by Eq.~(\ref{eq:analytical}) but only if the depth of the spherical shell is thin compared to its outer radius. Such a thin fluid layer is hardly justifiable for the solar convection zone. Notably, our model also unveils a branch of lightly damped, retrograde-propagating inertial modes with equatorially antisymmetric vorticity, whose dominant spherical harmonic components $T_{lm}(r)Y_l^m$ near the surface are such that $l=m+1$, and have frequencies close to the wave frequencies observed by \cite{Hanson+22}. All of these qualities match observations, leading us to an unequivocal identification of the HFR waves.

The modes we identify with the HFR waves belong to a class of inertial modes distinct from the Rossby mode class. Although their toroidal $(l=m+1,m)$ {component is much larger than the radial component} near the solar surface, their poloidal kinetic {energy is still comparable to the toroidal one, as opposed to Rossby modes.}

Our numerical calculations also suggest that the signature of low frequency tesseral-like Rossby modes with antisymmetric vorticity might be present in the observations presented by \cite{Hanson+22}. As true Rossby modes, their kinetic energy over the whole convective zone is mostly toroidal, in contrast with the modes in the main branch. Near the solar surface the tesseral-like Rossby modes have spherical harmonic $(l,m)$ components with contributions from $l=m+1,\,m+3,\,m+5$, and higher orders to a lesser extent, which appears to match the observations as well.

Differential rotation {and density stratification are} perhaps the most important features missing in our model. However, they are not essential for the identification of the modes. Our aim here is to provide the initial identification of the modes as a starting step towards more refined models, hoping to spark interest from the community in developing inertial wave models involving differential rotation, magnetic fields, and other effects. Such models, although numerically challenging, are in principle straightforward to develop. More refined inertial wave models \citep[e.g.][]{bekki2022} are potentially very valuable to understand and characterize turbulent processes in the solar convective zone  \citep{gizon2021}, but just as well in the convective zones of other stars where oscillations in the inertial range have been detected \citep[e.g.][]{ouazzani2020}. Inertial-wave-based helio/asteroseismology has great potential and might give us new insights into the interior dynamics of the Sun and the stars.

\begin{acknowledgments}
S.~T. and J.~R. express their warm gratitude to Veronique Dehant and Tim Van Hoolst for their encouragement and support. They also acknowledge financial support from the European Research Council (ERC) under the European Union's Horizon 2020 research and innovation program (Synergy Grant agreement no. 855677 GRACEFUL). A.~B. would like to thank Sabine Stanley for her encouragement and support. J.~R. would like to thank Bruce Buffett and UC Berkeley's Department of Earth and Planetary Science for their hospitality during the writing of this letter. We would like to thank as well Y. Bekki, K. R. Sreenivasan, and two anonymous reviewers for their constructive and insightful comments which helped us immensely to improve this work. 
\end{acknowledgments}

\vspace{5mm}


\software{\texttt{kore} \citep{triana2019,rekier2019}, SHTns \citep{shtns}, PETSc/SLEPc \citep{petsc-user-ref,petsc-efficient, Dalcin2011,slepc-toms,slepc-manual}, MUMPS \citep{MUMPS01,MUMPS02}, Matplotlib \citep{matplotlib}.
We provide \texttt{kore} as an open source and freely available code via \url{https://doi.org/10.5281/zenodo.6783310}. All the data and scripts needed to reproduce all the figures in this work can be found at \url{https://doi.org/10.5281/zenodo.6787680}.}

\appendix

\section{Numerical method} \label{sec:apx}
We expand the poloidal $P_{lm}(r)$ and toroidal $T_{lm}(r)$ functions occurring in Eq.~(\ref{eq:u0}) using a Chebyshev polynomial basis. We write
\begin{equation}
    P_{lm}(r) =\sum_{k=0}^N \alpha_{lm}^k\,t_k(x),\quad
    T_{lm}(r) =\sum_{k=0}^N \beta_{lm}^k\,t_k(x),
\end{equation}
where $t_k(x)$ is the Chebyshev polynomial of degree $k$, $N$ is the radial truncation level, the radial variable is mapped to $x$ via the affine transformation
\begin{equation}
    x=2\frac{r-r_c}{R_\odot-r_c}-1,
\end{equation}
and $\alpha_{lm}^k,\,\beta_{lm}^k$ are the unknown coefficients. {This is essentially the same technique used by \cite{rieutord1997}, except that we use a fast spectral method devised by \cite{olver2013}. This method uses Gegenbauer polynomial bases to represent the radial derivatives of $t_k(x)$ resulting in sparse matrix operators, as opposed to the spectral collocation method where the operators representing radial derivatives are dense matrices.} The resulting matrices representing Eq.~(\ref{eq:NS}) are banded and sparse. We end up with a generalized eigenvalue problem of the form
\begin{equation}
\label{eq:gep}
\vb{A}\vb{x}=\lambda\vb{B}\vb{x},
\end{equation}
where $\vb{A}$ and $\vb{B}$ are sparse matrices, $\lambda=(\sigma-\mathrm{i}\omega)/\Omega_\odot$ is the eigenvalue, and $\vb{x}$ is the eigenvector comprised by the set of coefficients $\{\alpha_{lm}^k,\,\beta_{lm}^k\}$. We use a \emph{shift-and-invert} strategy to obtain solutions with an eigenvalue close to a given target. The truncation levels used in our calculations are typically $N=156$ and $l_\mathrm{max}=158$ for the radial and angular expansions, respectively.

{
\begin{figure}
\plottwo{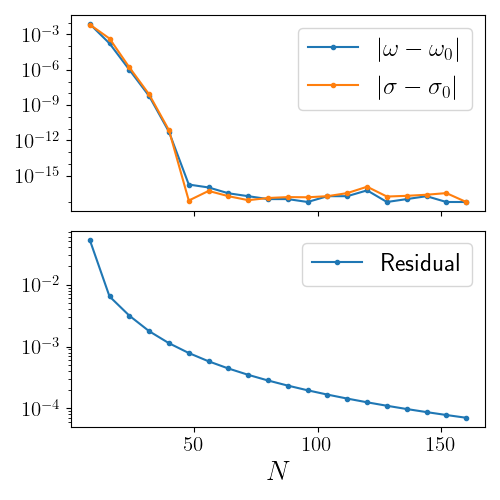}{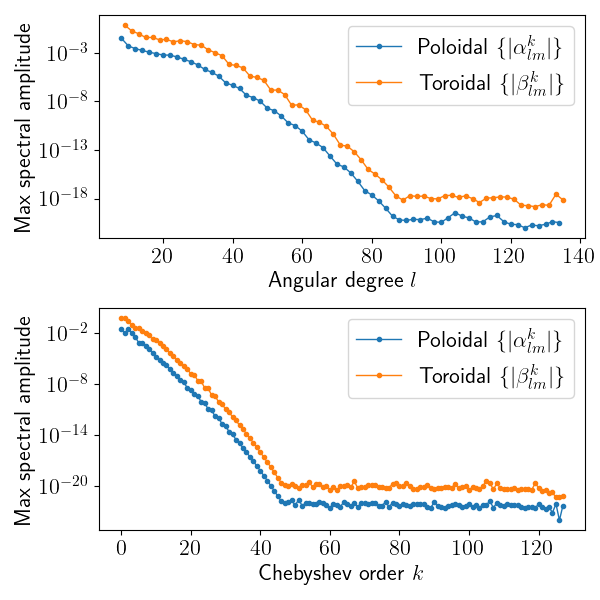}
\caption{On the top, left panel we show the effect of changing the truncation level $N$ on the eigenvalue of the $m=8$, $\omega/\Omega_\odot=-0.641$ eigenmode depicted in Fig.~\ref{fig:fig4} as a representative example. The reference eigenvalue $\sigma_0-\mathrm{i}\,\omega_0$ corresponds to the solution with N=160. The angular truncation $l_\mathrm{max}$ is linked to $N$ so that $l_\mathrm{max}\sim N$. Bottom, left panel shows the residual (defined in the text). The panels on the right show the maximum magnitude of the Chebyshev coefficients for a given $l$ (top), and for a given $k$ (bottom), associated with a particular solution with N=128, $l_\mathrm{max}=135$. 
\label{fig:converg}}
\end{figure}

Figure~\ref{fig:converg} gives an idea of the accuracy and convergence properties of the method we use. The top left panel shows the change on the eigenvalue as the truncation level is increased. Machine precision level on the variation is achieved around $N\sim50$. A measure of how well Eq.~(\ref{eq:NS}) is fulfilled is given by the residual plotted on the bottom left panel. We define the residual as
\begin{equation}
   \mathrm{Residual}= \frac{|2\sigma K-\mathcal{D}|}{\mathrm{max}\{|2\sigma K|,|\mathcal{D}|\}},
\end{equation}
where $K$ is the kinetic energy integrated over the whole fluid volume $V$:
\begin{equation}
K=\frac{1}{2}\int \vb{u}_0^\dagger\cdot\vb{u}_0\, \mathrm{d}V,
\end{equation}
and $\mathcal{D}$ is the viscous dissipation defined as \begin{equation}
\mathcal{D}=E\int\vb{u}_0^\dagger\cdot\nabla^2\vb{u}_0\, \mathrm{d}V.
\end{equation}
Our code was designed originally to study the flow in planetary interiors which typically have very small Ekman numbers and are thus computationally much more demanding compared to the Ekman number associated with the solar convection zone.
} Our code (named \texttt{kore}) is freely available as an open-source project: \url{https://doi.org/10.5281/zenodo.6783310}.
{
\section{Symmetry considerations} \label{sec:apx2}
}
The waves reported by \cite{Hanson+22} are described in terms of the radial vorticity observed near or at the solar surface. The spherical harmonic coefficients of the radial vorticity at any radius $r$ are related in a simple way to the toroidal functions $T_{lm}(r)$:
\begin{equation}
    \left[\vb{\hat r} \cdot (\nabla\times \vb{u}_0)\right]_{l,m} = l(l+1)\frac{T_{lm}(r)}{r}.
\end{equation}
The velocity field $\vb{u}$ of an inertial mode can be either equatorially symmetric or antisymmetric. The vorticity $\nabla\times\vb{u}$ has the \emph{opposite} equatorial symmetry as $\vb{u}$. Explicitly, if a mode is equatorially antisymmetric in the  vorticity, it fulfills
\begin{equation}
\begin{split}
    u_r(r,\pi-\theta,\phi) &= u_r(r,\theta,\phi),\\
    u_\theta(r,\pi-\theta,\phi) &= -u_\theta(r,\theta,\phi),\\
    u_\phi(r,\pi-\theta,\phi) &= u_\phi(r,\theta,\phi),
\end{split}
\end{equation}
i.e. it is equatorially symmetric in the velocity.
An inertial mode equatorially symmetric in velocity, with azimuthal wave number $m$, has poloidal functions $P_{lm}$ with indices such that $l=m,m+2,m+4,\ldots$, and the toroidal functions $T_{lm}$ have indices such that $l=m+1,m+3,\ldots$. Conversely when the mode is antisymmetric.

Inertial modes are a general class of modes that include Rossby modes (also known as \emph{planetary waves}) as a subset. The analytical dispersion relation for Rossby modes, namely
\begin{equation}
    \frac{\omega}{\Omega_\odot}=-\frac{2\,m}{l(l+1)},
    \label{eq:analytical}
\end{equation}
is usually derived assuming a thin spherical shell, which ignores motion in the radial direction, leaving only the toroidal part \citep[see e.g.][]{rieutord2014}. In the ideal case, a Rossby mode has only one spherical harmonic $Y_l^m$ component. These modes can then be classified as \emph{sectoral} if $l=m$, or \emph{tesseral} if $l>m$. A sectoral Rossby mode is by necessity equatorially antisymmetric in the velocity (i.e. symmetric in vorticity). These are the modes that have been observed and identified in the Sun by \cite{Loptien+18}. Tesseral Rossby modes can have either symmetry. Similarly, a Rossby mode with an equatorially symmetric velocity (i.e. antisymmetric in vorticity) is thus a tesseral Rossby mode. When the shell is not thin, poloidal motions can become significant, the dispersion relation (\ref{eq:analytical}) is not generally valid, and the foregoing classification does not follow unmodified. {Tesseral modes are the most affected, with other spherical harmonics appearing in the solution besides the main tesseral component, although the tesseral component remains dominant. Thus we call them \emph{tesseral-like}. Sectoral modes are essentially unaffected.}

{
\section{Differential rotation background} \label{sec:apx3}

Our model assumes a uniformly rotating background flow. However, we can obtain a crude estimate of the effect of the solar differential rotation background on the eigenfrequencies by considering the range of solar rotation frequencies corresponding to the spatial location (in the $r,\theta$ plane) where the amplitude of a given mode is substantial (averaged over $\phi$ and over an oscillation period). We define the `domain' of a mode as the region where its power is at least one half of its maximum power, or equivalently, where its amplitude is at least $\sqrt{2}/2$ times its maximum amplitude. We compute then the difference $\Delta\Omega=\Omega_\mathrm{max}-\Omega_\mathrm{min}$ between the maximum and minimum solar rotation rate \citep[based on][]{larson2018} \emph{within} the mode's domain. The shaded areas in Fig.~\ref{fig:fig1} represent $\omega \pm \Delta\Omega$, where $\omega$ is the mode's frequency from the uniformly rotating model.

Note that if a mode has substantial amplitude only in regions where the solar rotation rate can be considered as constant, say $\Omega_0$, then we expect no effect on the eigenfrequency, i.e. the uniform rotation model would be accurate, provided the frequency is referred to a reference frame rotating at $\Omega_0$. 
}
\bibliography{manuscript}{}

\begin{thebibliography}{}
\expandafter\ifx\csname natexlab\endcsname\relax\def\natexlab#1{#1}\fi
\providecommand{\url}[1]{\href{#1}{#1}}
\providecommand{\dodoi}[1]{doi:~\href{http://doi.org/#1}{\nolinkurl{#1}}}
\providecommand{\doeprint}[1]{\href{http://ascl.net/#1}{\nolinkurl{http://ascl.net/#1}}}
\providecommand{\doarXiv}[1]{\href{https://arxiv.org/abs/#1}{\nolinkurl{https://arxiv.org/abs/#1}}}

\bibitem[{{Abramenko} {et~al.}(2011){Abramenko}, {Carbone}, {Yurchyshyn},
  {Goode}, {Stein}, {Lepreti}, {Capparelli}, \& {Vecchio}}]{abramenko+11}
{Abramenko}, V.~I., {Carbone}, V., {Yurchyshyn}, V., {et~al.} 2011, \apj, 743,
  133, \dodoi{10.1088/0004-637X/743/2/133}

\bibitem[{Amestoy {et~al.}(2001)Amestoy, Duff, L'Excellent, \&
  Koster}]{MUMPS01}
Amestoy, P.~R., Duff, I.~S., L'Excellent, J.-Y., \& Koster, J. 2001, SIAM
  Journal on Matrix Analysis and Applications, 23, 15

\bibitem[{Amestoy {et~al.}(2006)Amestoy, Guermouche, L'Excellent, \&
  Pralet}]{MUMPS02}
Amestoy, P.~R., Guermouche, A., L'Excellent, J.-Y., \& Pralet, S. 2006,
  Parallel Computing, 32, 136

\bibitem[{Andersson(1998)}]{andersson1997}
Andersson, N. 1998, Astrophys. J., 502, 708, \dodoi{10.1086/305919}

\bibitem[{Balay {et~al.}(1997)Balay, Gropp, McInnes, \&
  Smith}]{petsc-efficient}
Balay, S., Gropp, W.~D., McInnes, L.~C., \& Smith, B.~F. 1997, in Modern
  Software Tools in Scientific Computing, ed. E.~Arge, A.~M. Bruaset, \& H.~P.
  Langtangen (Birkh{\"{a}}user Press), 163--202

\bibitem[{Balay {et~al.}(2019)Balay, Abhyankar, Adams, Brown, Brune,
  Buschelman, Dalcin, Eijkhout, Gropp, Karpeyev, Kaushik, Knepley, May,
  McInnes, Mills, Munson, Rupp, Sanan, Smith, Zampini, Zhang, \&
  Zhang}]{petsc-user-ref}
Balay, S., Abhyankar, S., Adams, M.~F., {et~al.} 2019, {PETS}c Users Manual,
  Tech. Rep. ANL-95/11 - Revision 3.11, Argonne National Laboratory

\bibitem[{Baruteau \& Rieutord(2013)}]{baruteau2013}
Baruteau, C., \& Rieutord, M. 2013, Journal of Fluid Mechanics, 719, 47

\bibitem[{{Baumann} {et~al.}(2004){Baumann}, {Schmitt}, {Sch{\"u}ssler}, \&
  {Solanki}}]{baumann+04}
{Baumann}, I., {Schmitt}, D., {Sch{\"u}ssler}, M., \& {Solanki}, S.~K. 2004,
  \aap, 426, 1075, \dodoi{10.1051/0004-6361:20048024}

\bibitem[{Bekki {et~al.}(2022)Bekki, Cameron, \& Gizon}]{bekki2022}
Bekki, Y., Cameron, R.~H., \& Gizon, L. 2022, arXiv preprint arXiv:2203.04442

\bibitem[{Christensen-Dalsgaard {et~al.}(1991)Christensen-Dalsgaard, Gough, \&
  Thompson}]{jcd1991}
Christensen-Dalsgaard, J., Gough, D., \& Thompson, M. 1991, The Astrophysical
  Journal, 378, 413

\bibitem[{Dalcin {et~al.}(2011)Dalcin, Paz, Kler, \& Cosimo}]{Dalcin2011}
Dalcin, L.~D., Paz, R.~R., Kler, P.~A., \& Cosimo, A. 2011, Advances in Water
  Resources, 34, 1124,
  \dodoi{http://dx.doi.org/10.1016/j.advwatres.2011.04.013}

\bibitem[{Gizon {et~al.}(2021)Gizon, Cameron, Bekki, Birch, Bogart, Brun,
  Damiani, Fournier, Hyest, Jain, {et~al.}}]{gizon2021}
Gizon, L., Cameron, R.~H., Bekki, Y., {et~al.} 2021, Astronomy \& Astrophysics,
  652, L6

\bibitem[{Greenspan(1968)}]{greenspan1968}
Greenspan, H. 1968, The Theory of Rotating Fluids, Cambridge Monographs on
  Mechanics (Cambridge University Press)

\bibitem[{Greer {et~al.}(2016)Greer, Hindman, \& Toomre}]{greer2016}
Greer, B.~J., Hindman, B.~W., \& Toomre, J. 2016, The Astrophysical Journal,
  824, 128

\bibitem[{Guenel {et~al.}(2016)Guenel, Baruteau, Mathis, \&
  Rieutord}]{guenel2016}
Guenel, M., Baruteau, C., Mathis, S., \& Rieutord, M. 2016, Astronomy \&
  Astrophysics, 589, A22

\bibitem[{{Hanson} {et~al.}(2022){Hanson}, {Hanasoge}, \&
  {Sreenivasan}}]{Hanson+22}
{Hanson}, C.~S., {Hanasoge}, S., \& {Sreenivasan}, K.~R. 2022, Nature
  Astronomy, \dodoi{10.1038/s41550-022-01632-z}

\bibitem[{{Hathaway} \& {Upton}(2021)}]{Hathaway+21}
{Hathaway}, D.~H., \& {Upton}, L.~A. 2021, \apj, 908, 160,
  \dodoi{10.3847/1538-4357/abcbfa}

\bibitem[{Hernandez {et~al.}(2005)Hernandez, Roman, \& Vidal}]{slepc-toms}
Hernandez, V., Roman, J.~E., \& Vidal, V. 2005, {ACM} Trans. Math. Software,
  31, 351, \dodoi{https://doi.org/10.1145/1089014.1089019}

\bibitem[{Hunter(2007)}]{matplotlib}
Hunter, J.~D. 2007, Computing in Science \& Engineering, 9, 90,
  \dodoi{10.1109/MCSE.2007.55}

\bibitem[{K{\"a}pyl{\"a} {et~al.}(2020)K{\"a}pyl{\"a}, Rheinhardt, Brandenburg,
  \& K{\"a}pyl{\"a}}]{kapyla2020}
K{\"a}pyl{\"a}, P.~J., Rheinhardt, M., Brandenburg, A., \& K{\"a}pyl{\"a},
  M.~J. 2020, Astronomy \& Astrophysics, 636, A93

\bibitem[{Larson \& Schou(2018)}]{larson2018}
Larson, T.~P., \& Schou, J. 2018, Solar physics, 293, 1

\bibitem[{Lockitch \& Friedman(1999)}]{lockitch1999}
Lockitch, K.~H., \& Friedman, J.~L. 1999, The Astrophysical Journal, 521, 764

\bibitem[{{L{\"o}ptien} {et~al.}(2018){L{\"o}ptien}, {Gizon}, {Birch}, {Schou},
  {Proxauf}, {Duvall}, {Bogart}, \& {Christensen}}]{Loptien+18}
{L{\"o}ptien}, B., {Gizon}, L., {Birch}, A.~C., {et~al.} 2018, Nature
  Astronomy, 2, 568, \dodoi{10.1038/s41550-018-0460-x}

\bibitem[{Michel \& Rivi{\`e}re(2011)}]{michel2011}
Michel, C., \& Rivi{\`e}re, G. 2011, Journal of the Atmospheric Sciences, 68,
  1730

\bibitem[{Olver \& Townsend(2013)}]{olver2013}
Olver, S., \& Townsend, A. 2013, SIAM Review, 55, 462

\bibitem[{{Ouazzani, R.-M.} {et~al.}(2020){Ouazzani, R.-M.}, {Ligni\`eres, F.},
  {Dupret, M.-A.}, {Salmon, S. J. A. J.}, {Ballot, J.}, {Christophe, S.}, \&
  {Takata, M.}}]{ouazzani2020}
{Ouazzani, R.-M.}, {Ligni\`eres, F.}, {Dupret, M.-A.}, {et~al.} 2020, A\&A,
  640, A49, \dodoi{10.1051/0004-6361/201936653}

\bibitem[{Rekier {et~al.}(2019)Rekier, Trinh, Triana, \& Dehant}]{rekier2019}
Rekier, J., Trinh, A., Triana, S., \& Dehant, V. 2019, Geophysical Journal
  International, 216, 777

\bibitem[{Rieutord(2014)}]{rieutord2014}
Rieutord, M. 2014, Fluid dynamics: an introduction (Springer)

\bibitem[{Rieutord {et~al.}(2001)Rieutord, Georgeot, \&
  Valdettaro}]{rieutord2001}
Rieutord, M., Georgeot, B., \& Valdettaro, L. 2001, Journal of Fluid Mechanics,
  435, 103

\bibitem[{Rieutord \& Valdettaro(1997)}]{rieutord1997}
Rieutord, M., \& Valdettaro, L. 1997, Journal of Fluid Mechanics, 341, 77

\bibitem[{Rincon {et~al.}(2017)Rincon, Roudier, Schekochihin, \&
  Rieutord}]{rincon2017}
Rincon, F., Roudier, T., Schekochihin, A., \& Rieutord, M. 2017, Astronomy \&
  Astrophysics, 599, A69

\bibitem[{Roman {et~al.}(2021)Roman, Campos, Dalcin, Romero, \&
  Tomas}]{slepc-manual}
Roman, J.~E., Campos, C., Dalcin, L., Romero, E., \& Tomas, A. 2021, {SLEPc}
  Users Manual, Tech. Rep. DSIC-II/24/02 - Revision 3.16, D. Sistemes
  Inform\`atics i Computaci\'o, Universitat Polit\`ecnica de Val\`encia

\bibitem[{Schaeffer(2013)}]{shtns}
Schaeffer, N. 2013, Geochemistry, Geophysics, Geosystems, 14, 751,
  \dodoi{10.1002/ggge.20071}

\bibitem[{Skoki{\'c} {et~al.}(2019)Skoki{\'c}, Braj{\v{s}}a, Sudar,
  Ru{\v{z}}djak, \& Saar}]{skokic2019}
Skoki{\'c}, I., Braj{\v{s}}a, R., Sudar, D., Ru{\v{z}}djak, D., \& Saar, S.
  2019, The Astrophysical Journal, 877, 142

\bibitem[{Triana {et~al.}(2021)Triana, Dumberry, C{\'e}bron, Vidal, Trinh,
  Gerick, \& Rekier}]{triana2021}
Triana, S.~A., Dumberry, M., C{\'e}bron, D., {et~al.} 2021, Surveys in
  Geophysics, 1

\bibitem[{Triana {et~al.}(2019)Triana, Rekier, Trinh, \& Dehant}]{triana2019}
Triana, S.~A., Rekier, J., Trinh, A., \& Dehant, V. 2019, Geophysical Journal
  International, 218, 1071

\bibitem[{Vasil {et~al.}(2021)Vasil, Julien, \& Featherstone}]{vasil2021}
Vasil, G.~M., Julien, K., \& Featherstone, N.~A. 2021, Proceedings of the
  National Academy of Sciences, 118

\bibitem[{Zhang {et~al.}(2004)Zhang, Liao, \& Earnshaw}]{zhang2004}
Zhang, K., Liao, X., \& Earnshaw, P. 2004, Journal of Fluid Mechanics, 504,
  1–40, \dodoi{10.1017/S0022112003007456}

\end{thebibliography}
\bibliographystyle{aasjournal}

\end{document}